\documentclass[12pt,fleqn]{article}
\usepackage{amsmath,amssymb}
\usepackage{graphicx}
\newcommand{\std}{\ensuremath{{\rlap{--}0}}\,  }

\begin {document}
\bibliographystyle{unsrt}
\title{ An energy interconversion principle applied in  reaction dynamics  for  the determination
of equilibrium standard states}
\author{Christopher G. Jesudason\footnote{Correspondence: Christopher G. Jesudason, M019,Chemical Physics Group,  Department of Chemistry, University of Malaya, 50603 Kuala Lumpur, Malaysia; Fax:03-79674193; Tel: 03-7967-4270 E-mail: jesu@um.edu.my}\\
{\normalsize Chemistry Department, University of Malaya,}\\ [1 mm]
{\normalsize 50603 Kuala Lumpur, Malaysia}\\
}
\date{{\normalsize 28 May, 2005} }
\parindent=0mm
\maketitle
\newtheorem{thm}{Theorem}

\newtheorem{cor}{Corollary}

\newtheorem{dec}{Deduction}
\newtheorem{pos}{Postulate}
\newtheorem{con} {Conjecture}
\newtheorem{defn}{Definition}
\newtheorem{hypo}{Hypothesis}

\newcommand{\oq}{\textquoteleft}
\newcommand{\cq}{\textquoteright}

\abstract{\noindent Chemical and other species reaction theories involving thermodynamical equilibrium states characterized by a temperature parameter invariably utilize statistical mechanical equilibrium density distributions. Here, a definition of heat-work transformation termed  thermo mechanical coherence is first made, and it is conjectured that most molecular bonds have the above heat-work transformation property, which models a chemical bond as a "`centrifugal heat engine"' .  Expressions are derived for the standard Gibbs free energy, enthalpy, and entropy where the bond coordinates need not conform to a non degenerate Boltzmann state, since bond breakdown and formation are processes that have direction, whereas equilibrium distributions are derived when the Hamiltonian is of fixed form, which is not the case for chemical reactions using localized Hamiltonians. The empirically determined Gibbs free energy from a known molecular dynamics simulation of a dimer reaction $
2{\rm A} \rightleftharpoons {\mathrm A}_{\mathrm 2} $,   accords rather well with the theoretical estimate. A relation connecting the rate of reaction with the equilibrium constant and other kinetic parameters is derived and could place  the commonly observed linear relationship between the logarithms of the rate constant  and equilibrium constant on a firmer theoretical footing. These relationships  could include analogues of the Hammett correlations used extensively in physical organic chemistry, as well as others which are temperature dependent.  One prediction of the principles developed here is that the equilibrium standard reaction free energy is more dependent on the height of the intermolecular potential than its depth, so that the sign of the $\Delta G ^{\std}$ can change for varying barrier height with fixed well depth, which may appear  counter-intuitive. All the above developments  can be tested directly in simulations and therefore provides a fertile ground for further research with significant implications on how standard states are determined in relation to the direction of chemical reaction.This work treats the molecular bond using standard thermodynamics as if it were a system,  and it is anticipated that with the advent of single-molecule science and experiment, that might be one way in which molecular statistical thermodynamics would develop.    
  }
 
\section {\normalsize {INTRODUCTION,DEFINITIONS AND MODEL}}
\noindent 
The Bohr correspondence principle can be used in the standard sense to derive the quantum description from the mainly classical description given here by appropriate mapping of the classical variables to operators and their inner products (norms). The  standard states used  is described below. Normally, the standard state is considered  "`arbitrary"' and set to unity and  the activity coefficients are relative to this chosen concentration state $c_0$ (e.g. unit concentration or pressure, such as moles per litre, Bar or Atmosphere); two primary concepts are implied in such an arrangement: (a) the  assumption that the chemical potential expression $\sim kT\ln \frac{c_i }{(c_0 \gamma _0 )}$ refers to setting to zero the work of the  intermolecular forces (perfect fluid) relative to expansion from the reference state where the activity coefficient $\gamma_i=1$ and where all $\gamma$'s are dimensionless and only differences in work are detectable and (b) in other regimes, the isothermal work done is $\sim kT\ln \frac{c_1\gamma_1 }{(c_2 \gamma _2 )}$ for the transition between two states at concentrations $c_1$ and $c_2$ respectively. The $c_0 \gamma _0 $ term is absorbed in the standard chemical potential $\mu_i^{\rlap{--}0}$ where $\mu_i=\mu_i^{{\rlap{--}0}} + kT\ln\frac{\gamma_ic_i}{[1.0]}$. The absorption is interpreted as follows where the activity $a_i$ of a component is $a_i=c_i\gamma_i$ for systems where the standard state is at zero density, leading to an apparent singularity. The chosen standard state (which we define and provide values for) used here  $\mu_i^{\rlap{--}0}(T)$ will be at infinite dilution or zero concentration ($c_i\rightarrow0$). For the species (chemical) reaction 
$\sum\nolimits_{i = 1}^N {\nu _i } A_i  = 0 $ the standard free energy $\Delta G ^{\rlap{--}0}(T) $is $\Delta G ^{\rlap{--}0}(T)= \sum\nolimits_{i = 1}^N {\nu _i } \mu_i^{\rlap{--}0}(T)  \neq 0 $ cannot be functions of density or any other variable other than temperature, where the conventional chemical potential for the work done from this zero concentration state to the one at $c_i$ is 

\begin{equation}  \label{e1}
\mu_i=\mu_i^{{\rlap{--}0}} + kT\ln\frac{\gamma_ic_i}{\delta c_i} 
\end{equation}

where $ \delta c_i\rightarrow0 $ for the total isothermal work done for  the standard state at $c_i=0$ by simple integration of the work given by  $kT\int \frac{d(a_i)}{c_i\gamma_i}$. Clearly eqn.(\ref{e1}) is not uniquely defined for the present and tends to a singularity. We can rescale the potential and write 
\begin{equation}  \label{e2}
\mu_i' ={\mu_i^{\rlap{--}0}}' +kT \ln \frac{c_i\gamma_i}{(\delta c_iN)}={\mu_i^{\rlap{--}0}}'(T) +kT\ln\frac{c_i\gamma_i}{[1 {\rm Unit}]}
\end{equation}
where $\delta c_i N=1$ for all $\delta c_i$.We choose $\delta c_i$ to be arbitrarily  small such that  $\gamma_i \rightarrow 1$ as $c_i \rightarrow 0$. This limit accords with the Debye-Huckel theory of electrolytes, and in most other conventional descriptions. Then 

\begin{equation}  \label{e3}
\mu_i'(T)={\mu_i^{\rlap{--}0}}'(T) +kT\ln \frac{c_i\gamma_i}{[1 {\rm Unit}]}
\end{equation}
Setting the $\mu_i$ of (\ref{e1}) to be the same as $\mu'_i$ of \ref{e3} ($\mu_i=\mu'_i$)$\Rightarrow$ 
\[ \mu_i^{{\rlap{--}0}} + kT\ln\frac{\gamma_ic_i}{\delta c_i} = \mu_i^{{\rlap{--}0}} 
+kT\ln N_i +kT\ln \frac{c_i\gamma_i}{\delta c_i N_i}     \]

or

\[ \mu_i^\prime(T)=\mu_i^{{\rlap{--}0}} 
+kT\ln N_i +kT\ln \frac{c_i\gamma_i}{[1]}     \]
leading to 
\begin{equation}  \label{e4}
\mu_i^{\rlap{--}0\,\prime}(T)=\mu_i^{\rlap{--}0}(T)+T.Q_i
\end{equation}
where $Q_i=kT\ln N_i$ is an arbitrarily large number; the subscript $i$ refers to the "`base"' material state for our development of reactions founded upon  a basic unit (e.g. a proton or an elementary particle from which all other material states are referred to ). For instance, for the reaction $n{\rm A} \rightleftharpoons {\mathrm A}_{\mathrm n} $ the standard state $\mu_{A_n}^{\rlap{--}0\,\prime}(T)$ for $\delta c_i\rightarrow0$ must be of the form  
\begin{equation}  \label{e5}
\mu_{A_n}^{{\rlap{--}0}\,\prime}(T)=(nTQ_{A_n}) +n\mu_{A}^{{\rlap{--}0}}(T)+\delta_{A_n}^{{\rlap{--}0}}(T)
\end{equation}
where $\delta_{A_n}^{{\rlap{--}0}}(T)$ is the Gibbs Free energy of Formation relative to state $i$; explicit expressions will be provided for  the computation of $\delta_{A_n}^{{\rlap{--}0}}(T)$ for $n=2$ . Further define $(n\mu_A^{{\rlap{--}0}}+\delta_{A_n}^{{\rlap{--}0}}(T) )=\mu_{A_n}^{{\rlap{--}0}}(T)$ so that $\mu_{A_n}^{{\rlap{--}0}\,\prime}(T)=\mu_{A_n}^{{\rlap{--}0}}(T) +nTQ_{A_n}$
The above therefore is the form of the activity coefficients $\gamma$ used here. The other important presupposition used here and also implicitly in the literature  that must be emphasized  is that in writing down this form of the chemical potential, only problems about thermodynamical equilibrium are solved. It is  an assumption to extend these forms to nonequilibrium regimes. From the Gibbs criterion,it follows that the standard state of the chemical potential
$\mu^{\rlap{--}0}$ for the potential $\mu_j= \mu_j^{\rlap{--}0}(T,c\rightarrow0)+\ln\gamma_jc_{j\neq i}$ is such that 
\begin{equation}  \label{e6}
-\Delta G^{\rlap{--}0}=-\sum_{j=1}^M \mu_j^{\rlap{--}0}\nu_j = kT\ln K_e
\end{equation}
so that $\mu_j^\std(T,c\rightarrow0)$ may be chosen so that (\ref{e6}) obtains, where $K_e$ always refers to the equilibrium constant. Since $c\rightarrow 0$ is a limit and a value, the vant' Hoff equation also follows from such a limit due to cancellation of the specific entropies \cite[p.182-183]{ira} leading to the equation
\begin{equation}  \label{e7}
\frac{d\ln K_e}{dT}=\frac{\Delta H^\std}{kT^2}.
\end{equation}
From the Gibbs criterion,we have 
\begin{equation}  \label{e8}
-\Delta G^\std =-\sum_{i}^{M} \nu_i\mu_i^{\std } =-\sum_{i}^{M} \nu_i\mu_i^{\std \prime}=\ln K_e
\end{equation}
so that $\mu_{A_n}^{\std \prime}$ has a recurring term $nTQ$ which cancels out in equilibrium thermodynamics; we therefore by convention ignore this contribution by writing the chemical potential for substance $i$ as 
\begin{equation}  \label{e9}
\mu_i^{\prime\prime}=\mu_i^\std (T) + kT\ln \frac{c_i\gamma_i}{1.}
\end{equation}
From \ref{e8}, 
\[-\Delta G=0=\sum_{i}^{M} \nu_i\mu_i^{\prime\prime} =\sum_{i}^{M} \nu_i\mu_i^{\prime}=\sum_{i}^{M} \nu_i\mu_i
\]
which implies that $-\Delta G^\std = \ln K_e = -\sum_{i}^{M} \nu_i\mu_i^{\std } $.  We note that 
\begin{equation}  \label{e10}
\sum_{i}^{M} \nu_i\mu_i^{\std\prime}=\sum_{i}^{M} \nu_i\mu_i^{\std}+\sum_{i}^{M}(n_iTQ)\nu_i 
=\sum_{i}^{M} \nu_i\mu_i^{\std}
\end{equation}
because $\sum_{i}^{M}(n_iTQ)\nu_i=0$
\begin{dec} \label{dc1}
Subject to the truth of Gibbs' equilibrium criterion, the standard chemical potential of species $i$ $\mu_i^\std$ is its free energy at zero density ( i.e. the free energy of a single particle entity coupled to a thermal well) and the chemical potential  $\mu_i(T,c_i)$ at any other state may be written $\mu_i(T,c_i)=\mu_i^\std(T)+kT\ln c_i\gamma_i $ .
\end{dec}

\begin{dec} \label{dc2}
From (\ref{e5}), the standard state of any other composite species relative to the base species may be computed
as  $\mu_{A_n}^\std=n\mu_{A}^\std(T)+  \delta_{A_n}^\std(T) $, taking into account the free energy change to that state.
\end{dec}

\begin{dec} \label{dc3}
Since the temperature parameter is specified, any species ${\rm A_i} $ has single species thermodynamics i.e. that of interactions with a thermal reservoir at  zero density,where the interactions are solely external forces between the reservoir and the particle, and the equilibrium thermodynamical properties are the time average values for as long as the particle exists, and the only exchange of energy between the molecule and the external  universe is via the heat reservoir.
\end{dec}

\begin{pos} \label{ps1}
The Gibbs' postulate that the ensemble average equals the time average obtains, so that the species energetics are determined by the average of its motion in the phase space of associated with it. Moreover, since these are all single particle species that react or interact, they need not assume Gaussian-type distributions for non-degenerate states for the energy terms associated with the reaction where there is an effective change of the Hamiltonian associated with the motions, as when a switching Hamiltonian is used. 
\end{pos}

\begin{defn}\label{df1}
A molecular species $\rm{A_n}$ (composed of $n$ elementary A units) is an entity whereby the intermolecular or internal forces can be unambiguously distinguished from the external forces. 
\end{defn}

\begin{dec} \label{dc4}
From Deduction \ref{dc3}, if the single species is thermalized, the energy exchange with the reservoir can only be the result of the external forces of the reservoir applied to the molecular species and vice-versa, and hence the energy distribution of the  system would conform to the average energies of the various types of external composite motions (i.e.  rotational and translational)  relative to the composite body  $\rm{A_n}$.
\end{dec}

\begin{dec} \label{dc5}
Since the species is thermalized at any instant, the energy equations combine at all times the dynamics due to the internal and external forces, and the stochastic average of any "`static"' thermodynamical state variable is the result of the application of Postulate (\ref{ps1}). Reacting species exist  for a finite lifetime during which the internal and external forces are distinguishable.
\end{dec}

From Deduction(\ref{dc5}), it is possible to couple stochastic and dynamical laws to determine the time evolution, e.g. conservation of angular momentum and energy equipartition.

\begin{pos}  \label{ps2}
Insofar as the single composite body (e.g. molecule, particle, etc.) is an 'external body' with respect to the thermostat, its thermal energy modes ( e.g. rotational and translational motiion about the centre-of-mass (CM)) are determined by the laws (quantum or classical) regulating these motions. 
\end{pos}

\begin{defn}  \label{df2}
A physical  or chemical species species which conforms to the above description for standard states, Deductions (\ref{dc1}-\ref{dc5})and Postulates (\ref{ps1}-\ref{ps2}) is thermomechanically coherent.
\end{defn}

 \section{\normalsize {DESCRIPTION OF A THERMOMECHANICALLY COHERENT REACTING PARTICLES AND MOLECULES}}
\begin{figure}[htbp]\label{f2}
\begin{center}
\includegraphics[width=11cm]{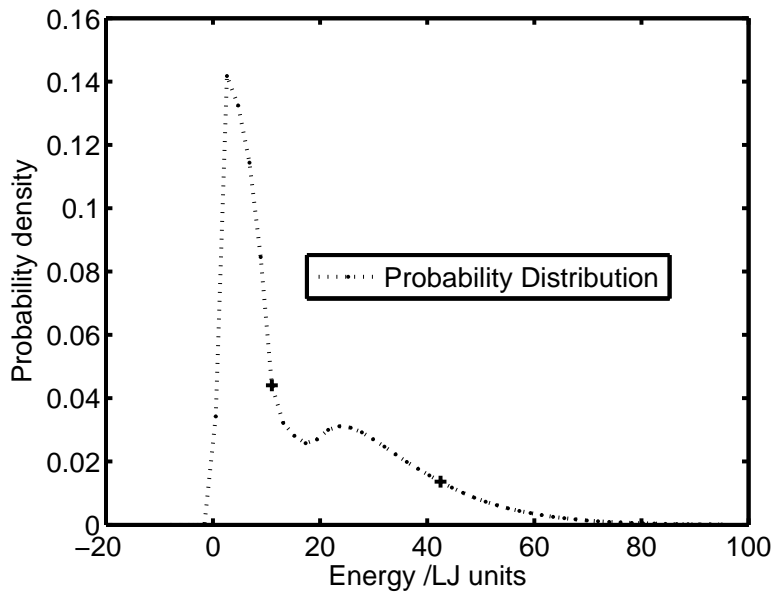}
\end{center}
\caption{Non-Boltzmann distribution of the total internal energy of the dimer in this equilibrium simulation}
\end{figure}

For the reaction 
\begin{equation}  \label{e11}
\rm{A_n}+\rm{A_m} \leftrightarrow \rm{A_{m+n}}
\end{equation}
a general scheme  is depicted in Fig.(\ref{f2}) with the appropriate coordinates. The simulation result reported here  is when $n=m=1$ for a simple dimer reaction. 
\begin{figure}[htbp]\label{f1}
\begin{center}
\includegraphics[width=11cm]{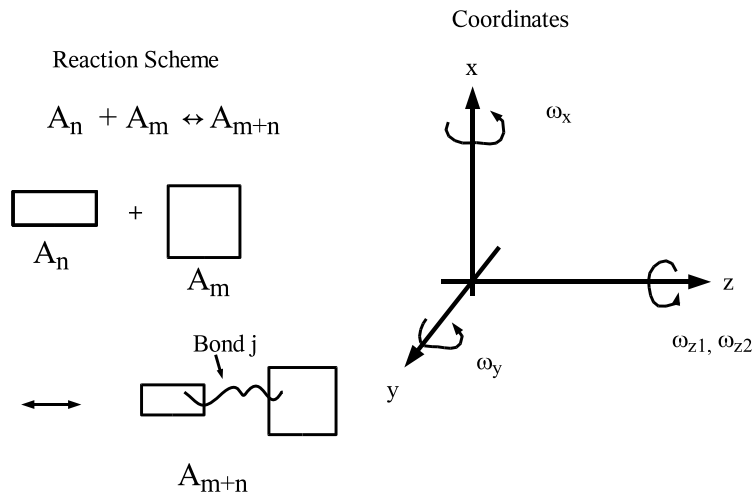}
\end{center}
\caption{Diagrams depicting the rotational axes and "`bond"' formation $j$  due to internal forces  }
\end{figure}
This  model is extended by considering product  formation about any bond $j$, depicted here along the $z$ coordinate. IF masses $A_n$ and $A_m$ were permanently bonded along $j$, then the Hamiltonian for the system would permanently feature the intermolecular potential energy (p.e.) between substances and the mean energy (for quantum systems) or equipartition of energy results from the density $P$ having form 
\begin{equation}  \label{e12}
P(\textbf{p,q})=e^{-\mathcal{H}(\textbf{p,q})}g(\textbf{p,q})
\end{equation}
 
 where $g$ represents the density in the total Hamiltonian space $(\textbf{p,q})$, which is not the same as the internal coordinates of the molecules, where if the  spatial coordinates $\textbf{R}\equiv\textbf{q}$ in the system subsumed that about the bond, then the vibrational motion would exhibit the equipartition result, but for a reacting system with bonds that break (approximating a switch of potentials) , the phase density is a function of the total system density and such equipartition results would not obtain. For instance, Fig.(\ref{f2}) shows the total internal energy $E_{int}$ distribution of the dimer in our simulation,where $E_{int}=\frac{1}{2}\mu \dot{r}^2+V(r)$ where $\mu$ is the reduced mass, $r$ the intermolecular distance, and $V(r)$ is the intermolecular potential (approximated as a Harmonic potential, where the end points which are curved in the actual potential was ignored). If there were no bond-beaks, or if they were very rare, then one would expect a Boltzmann-like distribution for the energy density, or the corresponding equilibrium densities given by QM, but this is not the case here since a real reaction is occurring where the internal coordinates are not the immediate phase space coordinates of the entire system Hamiltonian $\mathcal{H}$; for very small values of the rate constant, the masses at either end of the bond $j$ would be "`quasi-ergodic"'in that it would cover a fair portion of the phase-space of the internal coordinates, leading to a density $P$ given by the form (\ref{e12}) above.The internal energy is very approximately micro canonical in one arm of the distribution, $\mathcal{H}(T)=\overline{E}$ for the energy $E$ and temperature $T$ for the dimer studied here.This result should be contrasted with the many theories that presume that the reacting coordinate or internal coordinates have  Gaussian or equilibrium energy distributions e.g.  \cite{kurt1,aroeste1} or that the principle of local equilibrium obtains for these coordinates in the sense that Gaussian or equilibrium averaging is performed \cite{rubi1}. In the Eyring transition state theory, the vibrational coordinate connected to the "`bond"' does  not exist;the current mechanical description accords with this insight.   This result should be contrasted with advanced theories of bonding which all presume a Boltzmann or standard quantum probability  factor of the form $\exp-{\mathcal H}(\mathbf {p,q })/kT$ \cite{kurt1,aroeste1,jess1} for various subprocesses in kinetics. In \cite [p.3318]{kurt1}, the Maxwell-Boltzmann density is used to derive the polarization propagator in the binary collision approximation. The point here is that there is no reason to suppose this is necessarily true, since the relaxation times of other processes would determine whether or not such densities apply; for the total internal energy (vibrational energy) of the bond, Boltzmann-type densities might perhaps apply for relatively  very slow reactions  which have time to cover the phase space concerned for the potential of that configuration. The situation appears even more precarious in Aroeste's \cite[p.67]{aroeste1}development where "`...we now further assume an equilibrium distribution in internal states of the reactants..."' where in the simulation result here, the total internal energy of the bond does not conform to these assumptions. In a recent review \cite [eq. (1),p.68]{jess1}~of standard  Gibbs energy of association for large (protein) molecular fragments with ligands of form 
 \begin{equation}  \label{e13v2}
 A + B \Leftrightarrow AB
\end{equation}
  which is  of exactly  the same  form  as our dimer reaction, the standard reaction free energy is a complex function of Boltzmannized configurational integrals of form 
 \begin{equation}  \label{e13v3}
 Z_{N,A}= \int \exp [-\beta U(r_A,r_S)]dr_A dr_S
\end{equation}
 which includes the internal coordinates, where the complex "`contains six degrees of freedom that represent the residual translational and rotational motions of the bound ligand"'~with little vibrational coupling, since it is assumed that "`the effect of the ligand's translational/rotational motions on either species' internal vibrational motions are very small. "'. The standard Gibbs' free energy is then given by expressions of the form 
 \begin{equation}  \label{e13v4}
 \Delta G^\std_{AB}=-RT\ln \left(\frac{C^0}{8\pi^2}\right)\left(\frac{Z_{N,AB}Z_{N,O}}{Z_{N,A} Z_{N,B} }\right)
 + P^o\left\langle \Delta V_{AB}\right\rangle
\end{equation}
 where $R$ is the gas constant, $T$ the temperature, $C^0$ the standard concentration, the $Z$'s are the various configurational integrals for solute solvent and dimer interactions, and $P^o\left\langle \Delta V_{AB}\right\rangle$ is a pressure-work term. 
 In the model presented here, there is continuous coupling of the "`external"'  modes with the internal intermolecular forces. There seems to be a strong possibility  of elaborating  the method presented here in rudimentary form to such systems where further generalizations (and perhaps some corrections due to oversight at this preliminary stage ) to what is given here would  be required as more information and experimental facts are discovered. 
 The science of thermomechanics and continuum mechanics \cite{thermomech1} all use the standard heat and work terms of thermodynamics , together with various dissipation principles to couple processes and to describe  the non-equilibrium properties. The use of tensors and other mathematical tools are in a very advanced stage so that all kinds of cross-effects connected to the geometry  of the system (and symmetry) and coordinates of all the particles can be adequately accounted for. The work here focuses instead on the concepts of a new heat term and its connection to geometry, and doubtless the techniques currently available would be able to cast the concepts here into even more compact form. For instance, Lexcellent et. al  \cite{cth1} uses standard heat terms of equilibrium thermodynamical theory coupled with other terms  within a continuum; others concentrate on an amalgamation of basic linear thermodynamics description \cite[p.210]{cth2}, where  the laws of elasticity are derived  from the Gibbs potential~\cite[p.213,eqn.5]{cth2}. Others are able to couple motions \cite[p.184]{cth3}where the actual interior forces and the motions and power dissipated can be described  \cite[p.179]{cth3}but the difference  between mechanical energy and thermal energy in relation to  geometry and an external thermal source are not of immediate concern. 
 
 \subsection {\normalsize Description of dimer and coordinates}
 
In the simplest case, the molecule can change "`shape"', here defined in terms of the moments of inertia about the 3 principle axis, with the change described by three orthogonal coordinates in the direction of the 3 Orthogonal axes used to describe the molecule $r_{x,I},r_{y,I},r_{z,i,I}, (i=1....r)$ where  $\mathbf{r_I}=\left[r_{x,I},r_{y,I},r_{z,i,I}\right]$. Obviously, one must choose the coordinates convenient for the system of interest in the general case. As in Fig.(\ref{f1}), the "`bond"' is along the $z$ axis; if there is free rotation about the axis, then about bond $j$, $i\geq 2$ for  $r_{z,i,I}$ and the moments of inertia $I_j$ are functions of such changes of shape , i.e. for the simple case where there is no cross-coupling , we might write $I_j=I_j(r_{j,i}),\, (j=x,y,z;\; i \;\rm{for \; z\; coordinate\; only})$. These moments of inertia pass through the CM of the product species, where if $I_{i,j}=\sum_s m_s X_{i,s}X_{j,s}$\,, then $I_{i,j}=0 \; (i\neq j)$ where $i,j$ are the coordinates e.g.  $X_{x,s}=x_s,X_{y,s}=y_s$ ...etc. Concerning thermal energy, for  quantum theory, the high temperature rotational partition function has the form \cite[p.78]{mcc1} $f_{rot}$  where $f_{rot}=\left(\frac{8\pi^2kT}{h^2}\right)^{3/2}\left(\frac{\pi I_{xx}I_{yy}I_{zz}}{\sigma}\right)^{1/2}$ with the corresponding thermal energy $E_f$ given by
\begin{equation}  \label{e13}
E_f=kT^2\left(\frac{\partial\ln f}{\partial T}\right)_V,
\end{equation}
  i.e.  for classical and quantum theories, there exists a function $E_{thermal}=E_{TY}(T,I,\mathbf{r_I},TH)$ where $TY$ refers to theory type (either classical or quantum) and $TH$ always refers to the type of thermal energy (vibrational, translational etc.); for translational energy about the CM, $E_{trans}=\frac{kT}{2}$ for each orthogonal direction; the same results obtain quantum mechanically. For classical rotation about the $z$-axis (free rotation) we would have an energy of $(kT)/2$ per independent rotation axis, leading to $kT$ for the $z$-axis, for the free and independent rotation of the two reactant portions $\rm{A_n}$ and $\rm{A_m}$ about this axis. The situation is much more complicated for TY=quantum;  for free rotation, the partition function is of the form $f_{free}=\frac{\gamma}{\sigma_int}\left(I_{red}T\right)^{1/2},\;I_{red}=\left(\frac{I_nI_m}{I_m + I_m}\right)$ where $\gamma$ is a constant, and the partition function for internal rotations  with various activated states of angular rotation have the  form $f_{i.r.}=\sum_{i=0}^{\infty} \exp-(\omega_i -\omega_0)/(kT)$ but the exact form of Matthieu's equation is not known for the solution of the Schrodinger equation for hindered rotations \cite[p.89]{mcc1},and so several semi-empirical methods involving constants like $\gamma$  above  have been devised by Pitzer \cite{ptz1,ptz2} and others where the total partition function for rotation $f_{tot}$ can be written $ f_{tot}=f_{free}f_{i.r.}$ yielding the thermal energy from (\ref{e13}).   
The theory here is independent of TY; from the Bohr principal, one can start from classical considerations and from there derive the quantum expressions by applying the associated operators.
\subsection{\normalsize  Determination of work-heat energy balance with molecular distortion}
For rotational motion, consider an element of thermal energy due to rotation about the given ($z$) axis given by the form
\begin{equation}  \label{e14}
E_{TY}=E_{TY}(T,I_z,TH=\rm{rotn,z})
\end{equation}
Then  distortions to the molecule by the change in  $I_z$, in conjunction with  the application of the external thermal reservoir would cause a change of the thermal energy by amount $E_{TY}(T,I_z+\delta I_z,TH)$. On the other hand, prior to the application of the thermal reservoir interaction, the mechanical change of energy of rotation (with conservation of angular momentum when no external forces are acting on it) is $E_{\rm{Mech}}(I_z+\delta I_z,\rm{TH=rotn,z})$. This mechanical energy is derived directly from mechanics, (quantum or classical)  \emph{without statistical mechanics} since the body is a purely mechanical object at this interval of time. On the other hand, since the object must exist in a temperature field, and be in equilibrium with this thermal field and thermal reservoir, extra work of rotation $\delta w_{\rm{TH=rotn,z}}$ must be provided externally to the body to make it compatible with the thermal environment which would have the rotational energy $E_{TY}(T,I_z+\delta I_z,TH)$. Here, $I_z$ refers to the moment of inertia about the $z$ axis (i.e. the dimer is rotating in a plane perpendicular to the $z$-axis) There is conservation of angular momentum during the distortion, and the tangential impulses are applied by the reservoir to increase the kinetic energy (K.E.)to $E_{TY}(T,I_z+\delta I_z,TH)$.The analogy of this machine is the Carnot piston device, where if there is distortion of the cylinder, the $E_{TY}(TH=\rm{transl. \,\, and \,\,internal \,\, potential \,\, energy})$ thermal energy of the working substance is altered, and so the piston would have to be altered in position (by doing work) to maintain the same thermal energy density, where the work is $\delta w_p=-PdV$ ; (here $P$ is the pressure and $V$ the volume)and  thus we are lead to the following definition by the above analogy:
\begin{defn}  \label{df3}
An ideal centrifugal heat engine is defined as one where the work transfer increment $\delta w$ is the difference in energy due to the loss of rotational (mechanical ) energy, due
to the change in shape or distortion determined by the coordinates $\mathbf{R_z}$ and the prevailing thermal energy in a temperature field at $\mathbf{R_z+dR_z}$, where there is conservation of angular momentum during the distortion prior to the application of coupling moments orthogonal to the axis to restore the rotational energy to that of the rotational thermal energy at $\mathbf{R_z+dR_z}$.
\end{defn}
Likewise, we can define a rectilinear heat engine thus:
\begin{defn}  \label{df4}
An ideal rectilinear heat engine is one where losses in thermal energy along an external boundary coordinate during the application of an internal force by the system where there is total conservation of linear momentum can be compensated by the application  of an external force in the same direction, and where the work done by the external force is the difference in energy due to the loss of  thermal energy along the coordinate and the prevailing thermal energy at that same temperature. 
\end{defn}
 For rotational motion, the mechanical,non-thermal rotational energy (denoted $E_{Mech,\,Rot,z}(I(\mathbf{R_z}),\mathbf{\Omega})$ would suffer loss $\delta E_{Mech}$ if $\delta I_z >0$ because of conservation of angular momentum \cite[sec.11.6.1,p.253]{knu1} and the net work $dW$that would have to be done to enable energy balance for an ideal centrifugal heat engine is \begin{equation}  \label{e15}
d W {_z }  = \left( {dE_{TY} (TH = Rotn,z) - dE_{Mech,Rotn,z} } \right)
\end{equation}

For instance, if only the $z$ coordinates are involved then a simplified expression is
\begin{eqnarray}  
dW_{G,z}  &=& \left( \frac {\partial E_{TY} (TH = Rotn,z)}
{\partial \mathbf{R}_\mathbf{z} } \cdot \mathbf{dR}_\mathbf{z}\right. \nonumber \\
  & & \mbox{} - \left. \frac{\partial E_{Mech,Rotn,z} (TH = Rotn,z)}
{\partial \mathbf{R}_\mathbf{z} }\left|_{T.E.}  \right. \cdot \mathbf{dR}_\mathbf{z}  \right) =  - \mathbf{F}_{\mathbf{I,z}}  \cdot \mathbf{dR}_\mathbf{z} \label{e16}
\end{eqnarray}
where $T.E$ denotes a pathway along that of  thermal equilibrium, ,  $\mathbf{F_I,z}$ is the 
resultant internal force (due in part to the interaction with the thermal reservoir  so it is not the pure inter particle force that the molecule would experience in the absence of 
 thermal interaction) acting  along the coordinate $\mathbf{R_z}$ for the simple example where cross-coupling of forces has been eliminated. Eqn. (\ref{e16}) is a type of path differential, where the change
  $\frac{\partial E_{Mech}}{\partial \mathbf{R_z}}$  
  is along a thermal path that is consonant with equipartition of the thermal energy path at the specified temperature, as provided by quantum statistical mechanics. For simple systems, it can be shown by direct computation that the work transfer (heat absorption)  is at a maximum if (\ref{e15}) is utilized along each stage of the path $\mathbf{\delta R_z}$. It is suggested that non-rotational motion interconversion such as what obtains above for rotating systems is the essential method of transfer for work-heat balances in normal thermodyanical systems, which leads us to the following:
\begin{con} \label{conj1}
The rectilinear heat engine refers to the standard work-heat transformation of a fluid due to $P-V$ (Pressure-Volume) changes, such as is routinely discussed in thermodynamical treatises.
\end{con}

\begin{thm}\label{th1}

 The  change of the Gibbs energy between two states $\Delta G$ due to the transition between   two coordinate points    $\mathbf{R_z, R'_z}$ can be written 
\begin{equation}  \label{e17}
\Delta G_{\mathbf{ R_z}}^{\mathbf{ R'_z}}=G(\mathbf{ R_z})+\int_{\mathbf{ R_z}}^{\mathbf{ R'_z}}dW_{G,z}
\end{equation}  
\end{thm}
\textbf{Proof.}There is an interconversion of energy between the external thermal angular momentum K.E. and the internal energy along $\mathbf{ R_z}$ brought about by an external force so that there is no net heat transfer (pure work of Gibbs energy conversion at constant temperature). Hence (\ref{e16}) yields a change of the Gibbs energy between two states and the above results $\bullet$

\begin{hypo}  \label{h1}
The particle (including chemical) reactions found in nature are to a good approximation thermo-mechanically coherent if the reactions occur in an equilibrium temperature field.
\end{hypo}  

\section{\normalsize{TEST OF HYPOTHESIS}}\label{sec3}
In the following subsections , the model of a reaction is introduced,  which is applicable to all
elementary  second order reactions, and the general equations, derived   as theorems  are 
 applied to the above hypothesis  to  determine  the standard energy  states (standard Gibbs free energy, entropy and enthalpy of the  reaction). Estimates of the Gibbs energy is made based on the assumed shape of the molecular shape distribution function defined here; the determination of the function constitutes a challenging but solvable problem which will be pursued in forthcoming investigations, and the current theory can be verified from the said function. The estimates give results which are already in good quantitative   agreement with the Gibbs energy determined directly from the equilibrium constant.

\subsection{\normalsize The model}\label{subsec3.1}

The  dimeric  particle  reaction is 
\begin{equation}
2\text{A}\rightleftharpoons\text{A}_{2}  \label{e18}
\end{equation}
$(n,m=1)$ above  the supercritical
regime of the $LJ$ fluid. The model resembles that of ref. \cite{Gorecki3} except that
a harmonic potential is coupled to the products to form the bond of
the dimer whenever the internuclear distance reaches  the critical
value $r_f$ between two free atoms A and no virtual molecules are formed by labeling or coloring the product. Hence the internal states of the molecule can be probed, from which the standard states may be determined according to the theory developed here.
\begin{figure}[htbp] 
\begin{center}
 \includegraphics[width=11cm]{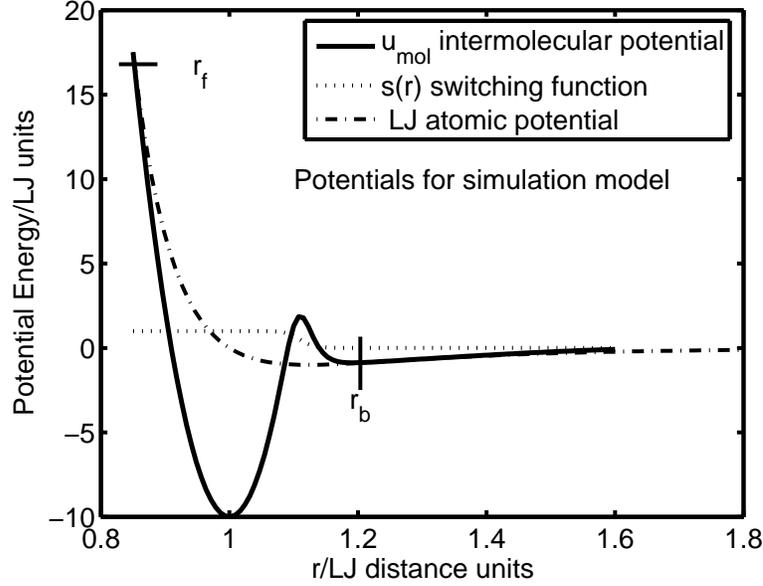}
 \caption{Potentials used for this work}\label{f3}
\end{center}
\end{figure}
In the current study, the potentials  as given  in Fig.~\ref{f3}
 are used, but other  configurations are possible. In this model, the dimer $\rm A_2$ is defined to exist the moment it passes the threshold at $r_f$ where the $u_{mol}$ intermolecular potential is activated.
The MD mechanism for bond formation and breakup is  as follows.
The free atoms A interact with
all  other particles (whether A or A$_{2}$) via
a Lennard-Jones spline potential and this type of potential has been
 described in great detail elsewhere \cite{Haf7}. An atom at a distance $r$ to another
particle possesses a mutual potential energy  $u_{LJ}$ where
\begin{align*}
u_{LJ}  & =4\varepsilon\left[  \left(  \frac{\sigma}{r}\right)  ^{12}-\left(
\frac{\sigma}{r}\right)  ^{6}\right]  \text{
\ \ \ \ \ \ \ \ \ \ \ \ \ \ \ \ \ \ \ for }r\leq r_{s}\\
u_{LJ}  & =a_{ij}(r-r_{c})^{2}+b_{ij}(r-r_{c})^{3}\text{
\ \ \ \ \ \ \ \ \ \ for }r_{s}\leq r\leq r_{c}\nonumber\\
u_{LJ}  & =0\text{ \ \ \ \ \ \ \ \ \ \ \ \ \ for }r> r_{c}\nonumber
\end{align*}
and 
\ where $r_{s}=(26/7)^{\frac{1}{6}}\sigma $ \cite{Haf7}. The
molecular cut-off radius $r_c$ of the spline potential is such that \ $r_{c}%
=(67/48)r_{s}$. The sum of particle diameters is $\sigma$\ and
$\varepsilon$ is the potential depth for interactions of type A-A 
or A-A$_{2}$. 
 The constants $a_{ij}$\ and $b_{ij}$\ where given before
\cite{Haf7} as 
\begin{align*}
a_{ij}  & =-(24192/3211)\varepsilon/r_{s}^{2}\nonumber\\
b_{ij}  & =-\left(  387072/61009\right)  \varepsilon/r_{s}^{3}%
\end{align*}
The potentials for this system  is illustrated in Fig.~\ref{f3}. 
 Any two unbounded atoms
interact with the above $u_{LJ}$ potential up to  distance  $r_{f}$
with energy $\varepsilon=u_{LJ}(r_{f})$ when the potential is switched
at the cross-over point  to the molecular potential given by

\begin{equation}
u_{mol}(r)=u_{vib}(r)s(r)+u_{LJ}\left[  1-s(r)\right]\label{e19}
\end{equation}
for the interaction potential  between the bonded particles   constituting the
molecule where $u_{vib}(r)$ is the vibrational potential given by
eq.(\ref{e19}) below and the switching function $s(r)$ has the form
given by eq.(\ref{e20}). At regions
$r<r_{sw},s(r)\rightarrow1$ implying $u(r)\sim u_{vib}(r)$,~i.e. the
internal force field is essentially harmonic for the molecule and at
distances $r<r_{sw},u(r)\sim u_{LJ}$, so that  the particle approaches
that of the free LJ type.Concerning the mechanism for the
switching,in quantum mechanical kinetic descriptions, switch
mechanisms are frequently used for describing potential crossovers\cite{Levine8},
but from a classical viewpoint  one can suggest that the inductive LJ
forces due to the particle potential field   (with particles
having a state characterized  by  state variables $\textbf{s}_{LJ} $)
causes the internal variables at the critical distances and
energies mentioned above to switch to state $\textbf{s}_{M} $ when
another force field is activated for the atoms of the dimer
pair. State  $\textbf{s}_{M}  $ reverts
again to state $\textbf{s}_{LJ} $ at distances $r_b$.  The following values of 
potential parameters were used  here (Fig.~\ref{f3}):
\newline
$u_0=-10,r_0=1.0,k\sim2446$ (exact value is determined by the other
input parameters),$n=100,r_f=0.85,r_b=1.20,\mbox{and}\; r_{sw}=1.11$.
\newline
The  intramolecular vibrational potential
$u_{vib}(r)$\  for a molecule is given by 
\begin{equation}
u_{vib}(r)=u_{0}+\frac{1}{2}k(r-r_{0})^{2}               \label{e19} %
\end{equation}
The  switching function $s(r)$ is defined as 
\begin{equation}
s(r)=\frac{1}{1+\left(  \frac{r}{r_{sw}}\right)  ^{n}}  \label{e20}
\end{equation}
where
\[\left\{\begin{array} {ll}
s(r)  & \rightarrow1\text{ \ \ \ \ \ if }r < r_{sw}\\
s(r)  & \rightarrow0\text{ \ \ \ \ \ for }r > r_{sw}
\end{array}
\right. \].
The switching function becomes effective when the distance between the
atoms  approach the value  $r_{sw}$ (see Fig.~(\ref{f3}) 
A molecule is formed when two colliding free 
particles have the potential  energy $u(r_{f})$ whenever  $r=r_{f}$
at the value  indicated  above.  This value  can be 
defined as  the isolated 2-body activation energy of the reaction. A molecule dissociates to  
two free atoms  when the internuclear  distance exceeds  $r_{b}$ (which in this case is 
1.20). Thus, the dimer $\text{A}_2$ exists the moment it passes the threshold at $r_f$ when the  $u_{mol}$ potential is activated, and the distribution of the dimer state is along the segments $a\equiv (0\leq r \leq r_f)$ and $b\equiv (r_f \leq r \leq r_b)$ leading to $c=a\cup b \equiv (0 \leq r \leq r_b)$.

LJ reduced units are used throughout this
 work unless stated otherwise by setting $\sigma ~\text{and}
 ~\varepsilon $ to unity in the above potential description.The
 relationship between normal laboratory units, that of the MD cell
 and the LJ units have been extensively tabulated and
 discussed\cite{Haf7}and will not be repeated here . For the system simulated here with the
potentials depicted in Fig.~(\ref{f3}), the switching function is operative
upto  $r_b$, the distance at which the molecule ceases to exist, and where
the atoms which were part of the molecule  interact with the potential
$u_{LJ}$ like other free atoms;bonded atoms interact with other
particles , whether  bonded or free with the $u_{LJ}$ potential. The
point $r_f$ of formation corresponds to the intersection of the
harmonic  $u_{vib}(r)$ and $u_{LJ}$ curves , and their gradients are
almost the same  at this point; by the Third dynamical law, momentum
is always conserved during the crossover despite finite changes in the
gradient. Total energy is conserved since the curves cross, and errors can
only be due to the finite time step per cycle in the Verlet leap frog
algorithm, which would cause the atoms to be defined as molecules at
distances $r<r_f$.Similarly at the point of breakup, there is a very
small ($\sim10^{-4}$ LJ units of energy) energy difference between the
LJ and molecular potentials despite using the switching function
 in the vicinity of the region to smoothen and unify the curves; the
 small energy differences at the cross-over points are less than that
 due to the normal potential cut-off at distance
 $r_c$ where the normal (unsplined) LJ potential is
 used in MD simulations. 
 A new algorithm which is general in scope was used to conserve momentum and energy  in this study and the thermostatting was at the ends of the MD cell, as would be the case in most real physical systems. If    $E_p (r)$ is the interparticle potential (energy) and  $E_m (r)$  that for the molecule just after the crossover, the algorithm promotes the particles to a molecule and rescales the particle velocities of only the two atoms forming the bond from $\mathbf{v}_\mathbf{i}$   to $\mathbf{v'}_\mathbf{i}\;\;(i=1,2)$      where 
 $\mathbf{v'}_\mathbf{i}  = (1 + \alpha )\mathbf{v}_\mathbf{i}  + \mathbf{\beta }$  such that energy and momentum is conserved, yielding  $\mathbf{\beta } = \frac{{ - \alpha (m_1 \mathbf{v}_\mathbf{1}  + m_2 \mathbf{v}_\mathbf{2} )}}
{{(m_1  + m_2 )}}$ (for momentum conservation) and energy conservation implies that $\alpha$  is determined from the quadratic equation $\alpha ^2 qa + 2qa\alpha  - \Delta  = 0$  with $a = (\mathbf{v}_\mathbf{1}  - \mathbf{v}_\mathbf{2} )^2 $ ,$q = \frac{{m_1 m_2 }}{{2(m_1  + m_2 )}}$  and $\Delta  = (E_p  - E_m )$  where empirically, success in real solutions for $\alpha$  for each instance of molecular formation is 99.9 \%  and 100\% for breakdown-where the $\Delta$  value in this   instance is very small ( $ \sim 1.0\times 10^{-4}$). This new algorithm coupled with shorter time step ensured excellent thermostatting. 
   The $y$ and $z$ directions of the MD cell
have length $1/16$ each (cell units). This shape is chosen because  it
is intended that future  simulations will 
concentrate on imposing thermal and flux gradients along the
$x-$axis,which would allow for more accurate sampling of steady state
properties about this axis. 
 conversion units
 The shape of the
potentials and switching mechanism used here is surprisingly similar
to discussions of the charge neutralization reaction given in \cite{Levine8}  
\begin{equation}
 \text{K}^+ +  \text{I}^- \rightarrow \text{K} +  \text{I} 
\end{equation}
except that  these  discussions do not explicitly mention the crossing
over of the $\text{KI}$ and $\text{K}^+\text{I}^-$ potentials at  
short distances (high energy),
although there is reason to suppose that such processes may well
occur. It seems very feasible that reactions that have an electro-magnetic force law component in the transition state may well exhibit some form of loop pathway such as the model used here.  
\subsection{\normalsize Energy distributions}\label{subsec3.2} 
Since this is an equilibrium system, Postulate (\ref{ps1}) leads to the dimer existing along a series of states about segment $c$ above. Defining $r_n=n\delta r_m$ where $\delta r_m$ is a fixed grid interval, we may determine the probability function by ensemble  averaging at low particle densities $(\rho^\ast=0.03\,\, \text{to}\,\, 0.08)$ by binning the number of occurrences $N_{r_n}$ at each of the intervals centered at $r_n=n\delta r_m$ so that 
\begin{equation}  \label{e21}
P_r(r=r_n)=\frac{N_{r_n}}{\sum_{r=0}^M N_{r_n}}
\end{equation}
where $M=\text{int}(r_b/\delta r_m)$. In the limit, $P_r(r)$ exists by Postulate(\ref{ps1}). As the dimer evolves about the r coordinate, work $dw$ is done on the dimer until the single molecule has a characteristic equilibrium temperature $T$; all dimer properties therefor, including the heat or work transfer characteristics must  be averaged by $P_r(r)$: by the Gibbs' postulate, the standard  free energy $\Delta G^\std (T)$ (determined at $\rho^\ast\rightarrow 0$), must also conform to the general deduction below for fixed temperature: $\delta G=\delta H -T \delta S$ and since $P=0$, $\delta G=\delta U -\delta q$. The total heat absorbed by the reservoir $-\delta q$ must be computed, as well as the total energy over the entire trajectory; both of these quantities are determinable from basic MD simulations.The thermodynamical path chosen on the single-particle micro system can be a work or heat exchange pathway, provided the same end-state is arrived at for both these paths since $G$ is a state function. We choose a work path in our example, rather than compute the quantities above.The height of the potential curve at $r_f$ is $17.5153$ and this is a work term.   From Definition (\ref{df1})  and Deduction (\ref{dc5}), we can view the particles and their aggregates as purely mechanical systems, thereby the "`bonds"' and other internal dynamics, without invoking any approximations whatsoever. the example of the dimer reaction assumes classical mechanics ($\text{TY=classical}$). The computations were over a range of temperatures $T^\ast = 4$ to $20$ so that  mean values of $\Delta G$ can be determined over this range (where   $T^\ast = kT$).The average value of $K_c=\frac{[A_2]}{[A]^2}$ ($[X]$ is the number density of species $X$) is determined for  densities in the range  $(\rho^\ast=0.03\,\, \text{to}\,\, 0.08)$ and this average was taken as an estimate of $K_e$, the equilibrium constant since $\gamma$, the activity coefficient $\rightarrow 1$; the fluctuations are considerable within this range and no discernible reduction in $K_c$ value could be detected.At $T^\ast=8$, each mode would have  from classical equipartition energy $T_f$, where $T_f=4$ (energy per mode at $r_f$. From these specifics, we can deduce that the thermo mechanical coherent thermal energy- work exchange  along a bond  between two internuclear distances  is given by   Theorem (\ref{th2})below:
\begin{thm}\label{th2}
The thermal energy-work exchange $\Delta W^{r_2}_{r_1}$ between internuclear distances $r_1$ and $r_2$ is given by 
\begin{equation}  \label{e24}
 \Delta W^{r_2}_{r_1}=-\left(\overline{E}_{rot}\right)\ln\left(\frac{I(r_1)}{I(r_2)}\right)
\end{equation}
\end{thm}
\textbf{Proof.}~For each of the $2$ rotational motions, $T_f=\frac{L^2_f}{2I_f}$, where $L,I$ and $T$ represents the angular momentum, moment of inertia and the total kinetic energy, with the subscripts indicating the position. When there is distortion of the $r$ coordinate, then $I$ varies, and the mechanical energy change in this case is $dE_{\text{Mech,Rot,z}}$ where 
\begin{equation}  \label{e22}
dE_{\text{Mech,Rot,z}}=-\frac{L^2(R_z)}{2}I^{-2}dI
\end{equation}
where $R_z=r$ the intermolecular coordinates. Since $T_f=\overline{E}_{rot}$ per mode, and $2IT_f=L^2$, leading to    $2I\overline{E}_{rot}=L^2$ so the increment of work $dW_G$  by substituting in (\ref{e22}) is given by  $dW_G=-I^{-1}\overline{E}_{rot}dE$ and since $\Delta W^{r_2}_{r_1}=\int_{r_1}^{r_2}dW_G$, the theorem follows $\bullet$

 Thus, there is a coupling of $I$ and $L$ because of the thermo-mechanical coherence property.  By Deduction(\ref{dc5}), we can compare result (\ref{e22})with the thermalization requirement of Equipartition, where from $2IT_f=L^2$ we have $2I4=L^2$, and from (\ref{e16}) of the equilibrium thermal path with known temperature $T$, we have
\begin{equation}  \label{e23}
dW_G=-4I^{-1}dI
\end{equation}
We note that   $\overline{E}_{rot}$ is the equipartition energy or energy per mode (either the quantum or non-quantum energy, depending on which theory is considered more suitable). Averaging leads to thermodynamical properties, such as the one below.
\begin{thm}\label{th3}
The total change of the  free energy about the  bond trajectory $\Delta G_{mol}$ is given by
\begin{equation}  \label{e25}
 \overline{W_{rf}}=\int_{r=0}^{r_f} \Delta W^{r_2}_{r_f}P(r_2) dr_2=\Delta G_{mol}.
\end{equation} 
\end{thm} 
 \textbf{Proof.}~Since all dimer states are equally valid, the  work done $\overline{W_{rf}}$  on the single dimer (for all members of the ensemble involved in this motion)  must be  (from (\ref{e24}) and (\ref{e21}) ) (per mode) the  time average or ensemble average value leading to the result above $\bullet$ 
 
The above  $\Delta G_{mol}$ is the increment of free energy associated with the bond in its trajectory.
In (\ref{e25}), for ant transition $\delta G = \delta H - T \delta S$ for any transition, so that if $\delta q$ is the input of rotational energy caused by internal stresses, then the internal work done is $\delta W^{In}$, and $\delta q$ is the heat absorbed where $\delta q=a$ and $\delta H =  \delta W + \delta q$, then $-T \delta S =- \delta q$, so that $\delta G = a+a-a=a= \delta W$.
Alternatively, instead of expansion, we can compress the single-molecular system, and (\ref{e25}) is derived if 
we note that if the bond axis about $z$ is compressed by doing work $\delta w$, so that the two other variations $\delta$ in the moments of inertia about the $x$ and $y$ principle axis obeys $\delta I_{x,y}<0$, then the increased kinetic energy of rotation would have to be removed by amount $-\delta q$ by the reservoir, where $\delta H=-\delta q + \delta w $ and since $-T \delta S =T\delta q$, then $\delta G= \delta w $, and integrating this leads to  $\Delta W^{r_2}_{r_f}=\int_{r_f}^{r_2}\delta w $ that can be inserted into (\ref{e25}). Finally some approximations may be derived.(Some exact expressions will be presented later. If the work done is approximately equal to the rotational energy dissipation, then $\delta U \approx  \delta w + \delta q =0$ or $\delta q \approx \delta w$, or $T\delta S \approx -\delta w$ or  by integration $T\Delta S_{r_f}^{r_2}\approx \Delta W ^{r_2}_{r_f}$
and so 
\begin{equation}  \label{e26}
T\left\langle \Delta S\right\rangle=T \int_{r=0}^{r_f}\Delta S_{r_f}^{r_2}P(r_2)dr_2
\approx \overline{W_{rf}}
\end{equation}
or 
\begin{equation}  \label{e27}
\left\langle \Delta S\right\rangle\approx\left(\frac{\overline{W_{rf}}}{T}\right)
\end{equation}

At present, the form of  $P$ is not available, since the simulation program 
developed did not anticipate the need for this distribution, but it will be included in future investigations for the years ahead, and would provide key evidence for or against  Hypothesis (\ref{h1})and the current formulation. But estimates of $P(r_2)$ can be made for extreme cases in the current study where  the real  value should lie in-between the extremes; this is what is found; further, the extreme cases  yield at the least semiquantitative results, meaning that quantitative results and corroboration of Hypothesis (\ref{h1}) is expected from a detailed study where $P(r_2)$ is known.
It will be noticed that the total bond energy $E_{tot}$ defined as $E_{tot}=U_m(r) + \frac{\mu}{2}\dot{r}^2$ is very approximately constant ($E_{tot}\approx c$) ,and  roughly simple Harmonic in nature, with the defining equation being
\begin{equation} \label{e28}
\mu \ddot r = {\bf F}_{{\bf int}} \quad \quad \left( {{\bf F}_{{\bf int}}  =  - \frac{{\partial U_m }}{{\partial r}}} \right).
\end{equation}

This motion would be moderated with the actual shape of the potential, and the collisions caused by the heat bath in the radial direction; such dynamics leads to the curve in Fig.(\ref{f2}). It is conjectured that the rate of the thermostatting might have an effect on the shape of the curve, as it would lead to  the boundary conditions of  (\ref{e28})  ~to be reset each time there is a collision.  The several scenarios which can occur are listed below:
\begin{itemize}  \label{item1}
\item [(1)] the molecule is formed at $r_f$ and then traverses to $r_b$ in such a way that  $P(r_2)=\text{constant}$ and $P(r<r_f)=0$. The motion approximates (\ref{e28}) except there is no accumulation of phase density at the turning points of the curve, so all points have equal weight (consistent with a transition stage with no multiple reflection back to the vicinity of $r_f$ with $E_{tot}=c$~). The array of conditions may be written 
\begin{eqnarray}
 I(r_1 ) &=& I(r_f ) = I_{r_f }, \nonumber \\ 
 I(r_2 ) &=& I(r_b ) = I_{r_b }, \nonumber  \\ 
 \overline{W_{r_f }} ^{(1)}  &=&  - 2(\overline{ E}_{rot} )\ln \left( {\frac{{I_{r_f } }}{{I_{r_b } }}} \right) \label{e29}
 \end{eqnarray}
The  parameters are $r_f=0.85,r_b=1.20,I_f=0.36125, I_b=0.720$. For the $2$ rotational modes at $T^\ast = 8, \overline{E}_{rot} =4.0$ for classical systems, so that (\ref{e29}) yields $\overline{W_{r_f }} ^{(1)}  = 0.6862T^\ast $
\item [(2)]since $\Delta G^\std(T^\ast =8)$ from the simulations, we can adjust $I(r_2)$ ($I_{r_1} =I_{r_f}$  is fixed) so that there is exact coincidence at $\Delta G^\std(T^\ast =8)$; this occurs when $r_f=1.1539$ (instead of $1.20$) and $I(r_b)=0.6656$ where $r_f$ and $I_f$ are as in (1) above. The $\overline{W_{r_f}}$ function, denoted in this instance  $\overline{W_{r_f}}^{(2)}$ is from (\ref{e29})~$\overline{W_{r_f}}^{(2)}=0.61110T^\ast$
\item [(3)]here we choose $I_{rat}=-\ln \left(\frac{I_{r_f}}{I_{r_b}}\right)=0.50$, where
$ \left(\frac{I_{r_f}}{I_{r_b}}\right) = 0.60653$  and the $W_{r_f}$ function in this case is 
$W_{r_f}^{(3)}=0.50T^\ast$
\end{itemize}
\subsection{\normalsize Derivation of standard thermodynamic functions for dimer reaction  }\label{subsec3.3} 

For the following transition scheme,

 \begin{eqnarray} \rm{ 2A} \rightarrow & \rm{A_2 \,(threshold\,\, at\,\, r_f=0.85)} \rightarrow & \rm{A_2 \,(dimer \,\,phase\,\, space)}\nonumber \\
\rm{State \,\,\,1 }&\rm{ State\,\, 1^\ast} & \rm{ State\,\, 2}\label{e30}
\end{eqnarray}
the molecule "`exists"' in the space ($\rm{State \,\, 1^\ast \cup\,\,State\,\,2}$).
\par
Notation: $P:{\rm Pressure};\,\, U,E:{\rm Energy\,\,function}\,\,;\, H:{\rm Enthalpy} \,\,;\,T,T^\ast:\rm{Temperature}$; asterisks imply reduced forms, e.g. $\epsilon ^\ast:=(kT)=T^\ast $; $\delta X$ :\,variation of $X$; $w$ denotes a work term, and $q$ always refer to heat variables; subscripts denote whether the dimer-molecule (m) or atomic species (a) is being referred to; L refers to the latent energy of the particles (which usually cancels out in the expressions). 

An elementary treatment of the statistics of a "`bond"' that can be dynamically broken and formed is given in Appendix (A.) \\
\underline{State $1\rightarrow1^\ast$ transition}\\
Since $P\rightarrow0$, $H=U$. So, 
\begin{equation}  \label{e31}
H_a=U_{particles}=\left(\frac{3}{2}\epsilon^\ast\times 2 +2E_L=
\frac{6}{2}\epsilon^\ast + 2E_L=\delta w_1 + \delta q_1 + 2E_L\right)
\end{equation}
 where $\delta w_1=0$ (no work). For molecules, we get
 \begin{eqnarray}  
 H_m &=& U_{molecule}=\delta W_m + 2E_L + q_{1^\ast} \nonumber \\
  &=& \frac{5}{2}\epsilon^{\ast}+E_L + \delta W_m \label{e32}
\end{eqnarray}
where $q_{1^\ast}=2\left( \frac{\epsilon^\ast}{2}\right) + \frac{3}{2} \epsilon^\ast $ for the translational and rotational energy of the composite structure. $E_L$ is the latent energy of the isolated atom at the indicated temperature. It is crucial to realize that $\delta W_m$ is the work done to "`fuse"' the particles at thermal equilibrium  into a single body (from the point of view of mechanics previously described)  and that after that process, the external reservoir only works upon a single external body that distorts itself, and the forces of interaction are  through  external forces; (\ref{e32}) represents the rotational and translational K.E. about the CM  and the residual energy which is carried over from the constituent reactant species and the  work $\delta W_m$that is required to fuse the reactants, where $\delta W_m = 17.5153$ and this comes from the p.e. at $r_f$ given in Fig.~(\ref{f3}).  The change in enthalpy   is therefore given by 
\begin{equation}  \label{e33}
\Delta H_{1 \rightarrow 1^\ast}=H_m -H_a=\delta W_m -\frac{\epsilon^\ast}{2}
\end{equation}
 and the entropy change $\Delta S$  is such that
 \begin{equation}  \label{e34}
 T\Delta S_{1 \rightarrow 1^\ast}=\frac{5}{2}\epsilon^\ast -  \frac{6}{2}\epsilon^\ast = -  \frac{1}{2}\epsilon^\ast
\end{equation} 

leading to 
\begin{equation}  \label{e35}
\Delta G_{1 \rightarrow 1^\ast}=17.5153 -  \frac{1}{2}\epsilon^\ast  -T\Delta S =17.5153 \,\,\,.
\end{equation}
A heuristic interpretation of the "`bond"' free energy is possible. The $z$ axis translational k.e. is $2.(\epsilon^\ast/2)$ and when the bond is formed, one can still refer to the heat content of the bond as $q_{bond}=\epsilon^\ast$. Since $\delta H = \xi_{max} + \epsilon^\ast$ and  $-T\delta S=-\epsilon^\ast$, the bond free energy change is $\Delta G^\ast_{bond}=(\xi_{max} + \epsilon^\ast)-\epsilon^\ast=\xi_{max}$.
\underline{State $1^\ast \rightarrow 2$ transition}\\
Thus, for thermo-mechanical coherent systems, as defined in (\ref{e25}), we have
\begin{equation} \nonumber 
\Delta G^\std (T) = \overline {\Delta G}(T) = \Delta G_{1 \rightarrow 1^\ast} + \int_0^{r_b}\left(\int_{1^\ast}^
{r_2}dG\right)P(r_2) dr_2
\end{equation}
or 
\begin{equation}  \label{e36}
\Delta G^\std (T) =\Delta G_{1 \rightarrow 1^\ast}+ \overline{W_{rf}}
\end{equation}
with $\overline{W_{rf}}$  defined as in (\ref{e25}). In general, however, 
\begin{equation}  \label{e37}
\Delta G^\std (T) =  \Delta G_{1 \rightarrow 1^\ast} + \int_0^{r_b}\left(\int_{1^\ast}^
{r_2}\left( dH-TdS\right)\right)P(r_2) dr_2
\end{equation}
. Eqn(\ref{e36})will  be considered in detail in the next section. We also define the following energy term
for the two cases that we may encounter:
\begin{equation}  \label{e37v2}
\overline{W_{rf}}~'=\left\{ \begin{array}{c}
\overline{W_{rf}}\,\,\,\text {from}\,(\ref{e25})\,\, \text {if~system~is~thermomechanically~coherent},\\
\text{otherwise~we~set}\\
\int_0^{r_b}\left(\int_{1^\ast}^
{r_2}\left( dH-TdS\right)\right)P(r_2) dr_2 \end{array} \right. .
\end{equation}

\subsection {Estimates of $\Delta G^\std (T)$}\label{subsec3.4}
From (\ref{e36}), and (\ref{e29}),we  derive 
\begin{eqnarray}
\Delta G^{\std (1)} (T) &=& \Delta G_{1 \rightarrow 1^\ast}+ \overline{W_{rf}}^{(1)}	= 17.5153 + 0.6862T^{\ast} \label{e38}\\
\Delta G^{\std (2)} (T) &=& \Delta G_{1 \rightarrow 1^\ast}+ \overline{W_{rf}}^{(2)}	= 17.5153 + 0.61110T^{\ast}
\label{e39}\\
\Delta G^{\std (3)} (T) &=& \Delta G_{1 \rightarrow 1^\ast}+ \overline{W_{rf}}^{(3)}	= 17.5153 +  0.500T^{\ast}\label{e40}
\end{eqnarray}

\begin{figure}[htbp] 
\begin{center}
 \includegraphics[width=11cm]{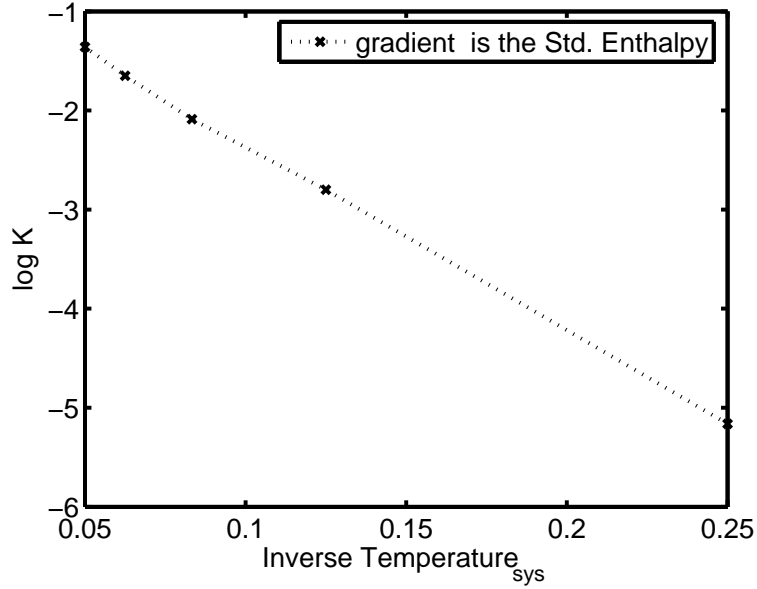}
 \caption{Linear Plot of $\ln K_e \,\,vs\,\, 1/T^\ast$}\label{f4}
 \end{center}
\end{figure}
The results from the simulation  for $\Delta S^\std ,\Delta H^\std \,\text{and}\,\Delta G^\std $  follows. Here, we can only discuss estimates for $\Delta G^\std $ based on the theory; the derivation for the enthalpy and free energy of reaction will be presented but no detailed discussion is useful at this stage until the $P(r_2)$ function is determined numerically so that a comparison between the theoretical predictions and the numerical values derived from simulation is made; this is a long term challenge. The van't Hoff equation (at constant pressure, $P \rightarrow 0$) is
 \begin{equation}  \label{e41}
 \frac{d\ln K_e}{d(1/T)}= -\frac{\Delta _rH^\std (T)}{R}
\end{equation}
A plot of $\ln K_e$vs $1/T^\ast$ is given in Fig.~(\ref{f4}) where the plot is rather linear over the range of experimental temperature points; therefore the gradient would yield a fairly constant average  value applicable for the temperature range for  the standard enthalpy $\overline{\Delta H^\std}$.  The overbar refers to these mean values over the temperature range ($T^\ast =5-20$). Since $\Delta G^\std$ is known, the average standard entropy may be determined from $N$ readings as 
\begin{equation}  \label{e42}
\overline{\Delta S^\std}=\left(\sum_{i=1}^N\frac{\Delta G^\std (T_i)-\overline{\Delta H^\std}}{T_i}\right)
\end{equation} 
The theoretically less correct method is to do a linear fit to the equation (assuming that both the standard entropy and enthalpy is reasonably constant without prior justification) 
\begin{equation}  \label{e43}
\Delta G^\std=\Delta H^\std -T\Delta S^\std .
\end{equation}
The van't  Hoff (\textit{v.H}.) results using (\ref{e41}) are 
\begin{eqnarray}
\overline{\Delta H_{v.H.}^\std}&=& 18.8 \pm 0.6 \label{e44} \\
\overline{\Delta S_{v.H.}^\std} &=& -0.47 \pm 0.06  \label{e45} \\
{\Delta G_{v.H.}^\std}  &=& \overline{\Delta H_{v.H.}^\std} -T  \overline{\Delta S_{v.H.}^\std}. \label{e46}
\end{eqnarray}
At $T^\ast=8$, the experimentally determined $\Delta G^\std$  (from extrapolation to zero density) is
\begin{equation}  \label{e47}
\Delta G^\std=22.4 \pm 0.3
\end{equation}
The \textit{v.H} equation determination (\ref{e46}) for the same temperature  is 
\begin{equation}  \label{e48}
{\Delta G_{v.H.}^\std} =22.5 \pm 0.6
\end{equation}
which is close to the experimentally determined value. The linear fit (\textit{l}) estimate (\ref{e43}) for the
entropy and enthalpy are 
\begin{eqnarray}
\overline{\Delta H_{l}^\std}&=& 19.2 \pm 0.9 \label{e49} \\
\overline{\Delta S_{l}^\std} &=& -0.43 \pm 0.07  \label{e50v1} 
\end{eqnarray}
Surprisingly, perhaps, (\ref{e49},\,\ref{e50v1})are reasonably close to the \textit{v.H.} results. Furthermore, from estimate (\ref{e27}), we observe that $\overline{\Delta S_{l}^\std}$ is close in value to the range ($0.500\text{ --}\, 0.6862$) derived from (\ref{e38}-\ref{e40}).

\subsection{Interpretation of results}\label{subsec3.5}
\begin{figure}[htbp] 
\begin{center}
 \includegraphics[width=11cm]{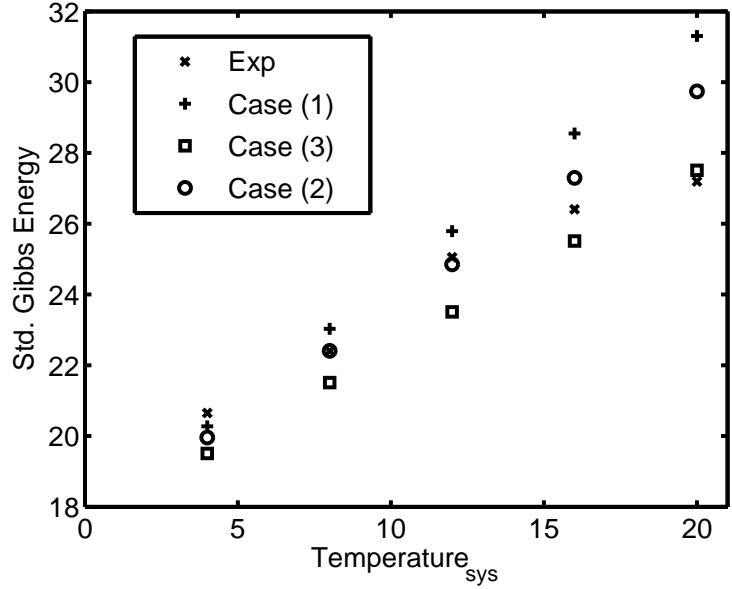}
 \caption{Plots of  the estimation of the  standard Gibbs' energy for the   three cases discussed in the text together with  the simulation results.}\label{f5}
 \end{center}
\end{figure}

The values ($\Delta G^{\std (1-3)}$) from (\ref{e38}-\ref{e40}) plotted against the experimental curve given in 
Fig.~(\ref{f5}) is 
essentially quantitative even for the approximations made here. $\Delta G^{\std (1)}$ (Case (1)) represents a distribution of "`dimer states"' uniformly distributed, and is quantitative for $4<T^\ast < 10$ (lower temperatures).The cases referred to in the superscripts have the following moments of inertia ratios
\begin{equation}  \label{e50}
\left(\frac{I_{r_f}}{I_{r_b}}\right)_1=0.502;\,\,\left(\frac{I_{r_f}}{I_{r_b}}\right)_2=0.543;\,\,
\left(\frac{I_{r_f}}{I_{r_b}}\right)_3=0.607 \,\,.
\end{equation}
Relative to the loop mechanism, for fixed $I_{r_f}$, the results clearly show that there exists an effective value of the moment of inertia  determined by $P(r_2)$ designated   $I_{r_{b_x}}$
where $I_{r_{b_x}} < I_{r_b,1}$ 
(the arabic numerals refer to the Cases (1-3) )for optimal results if the theory is correct, and that this moment of inertia would change with temperature;  there is a non-uniform distribution of steady state distances with an accumulation of distance density at distances $r$ where $r<r_b$; i.e. there is a preponderance of states to the left of the region of dimer breakdown. Indeed, at high temperatures, the probability of states $x$ with internuclear distances $r_x\rightarrow r_f$ 
would increase in the vicinity $\delta$ about $r_f$ for arbitrary $\delta$ so that $I_{r_{b_x}} <I_{r_b,1}$ leading to the observations. 
However, the overall changes are not very dramatic. In particular, once the effective $I_{r_f}/I_{r_b}$ratio has been selected to coincide with an experimental point (Case 2 at $T^\ast=8.0$) by adjustment of the parameter to fit with the actual $P(r_2)$ distribution function according to the prescription below (where the notation for $W$ is as in (\ref{e25})\,\,)   where 
\begin {equation}\label{e51}
-T^\ast \ln \left(\frac{I_f}{I_b}\right)_{fit} = \int_{r=0}^{r_b} \Delta W^{r_2}_{r_f}  P(r_2) dr_2
\end{equation}
then there is quantitative agreement about the whole range for $T<15$.
We have also to derive  from eqns.(\ref{e44},\ref{e45}) the $\Delta H^\std$ and $\Delta S^\std$terms formally for this theory, which must follow from the internal dynamics of the molecule. For  theorems \ref{th4} and \ref{th5}, the term $\Delta W^{r_2}_{r_f}$ can refer to either that of  (\ref{e36}) OR to the integrated Gibbs energy up  to coordinate $r_2$ , that is 
\begin{equation}  \label{e53v2}
\Delta W^{r_2}_{r_f} =\left\{ \begin{array}{l}
\Delta W^{r_2}_{r_f} \,\,\,\text {from}\,(\ref{e24})\,\, \text {if~system~is~thermomechanically~coherent},\\
\text{otherwise~we~set}\\
\int_{1^\ast}^
{r_2}\left( dH-TdS\right) \end{array} \right. .
\end{equation}

\begin{thm} \label{th4}
The standard entropy is given by 
\begin{equation}\label{e54}
	\Delta S^\std	(T)=-\int_{r=0}^{r_b} \frac{\partial}{\partial T}\left(\Delta W^{r_2}_{r_f}P(r_2,T)\right)dr_2 
\end{equation}
\end{thm}
\textbf{Proof.}
The fundamental thermodynamical relationship\cite[p.167, Sec. 7.2]{Ait2} $\Delta S^\std=-\frac{\partial \Delta G ^\std}{\partial T}\left|_{P,\,\mathbf{n}}\right.$ may be applied to (\ref{e36})written 
\begin{equation}\label{e53}
\Delta G^\std	(T)=\xi_{max}+ \overline{W_{rf}}=\int_{r=0}^{r_b} \Delta W^{r_2}_{r_f} dr_2 P(r_2,T) 
+\xi_{max}
\end{equation}
and differentiating this equation with respect to $T$ leads to the resulting theorem $\bullet$

Equation (\ref{e54})is testable from the point of view of simulations, so that an independent and explicit determination of $\Delta S^\std	(T)$ is available, which can be correlated with the entropy derived directly from curve fitting to the $\Delta G^\std	(T)$ as detailed above in the previous sections. We note the elementary fact that entropic values are intimately connected to probabilities, and so (\ref{e54})is another example; indeed if $TdS=\sum_{i=0}^M\epsilon_i dn_i$ where $\epsilon_i$ are energy  levels \cite[p.642]{Ait2} and $dn_i$ are the changes in population then $dS\sim \sum_{i=0}^M(\epsilon_i dn_i)/T$ which is of the same form approximately as (\ref{e54}). As with theorem (\ref{th1}) the standard enthalpy can also be derived.
\begin {thm} \label{th5}
The standard enthalpy of reaction $\Delta H^\std	(T)$ is given by 
\begin{eqnarray}
	\Delta H^\std	(T)&=&\xi_{max}+  \int_{r=0}^{r_b}\left(\Delta W^{r_2}_{r_f} \left\{ P(r_2,T) -T\frac{\partial P(r_2,T)}{\partial T}\right\}\right.\nonumber \\
	&-&\left.P(r_2,T)T\frac{\partial \Delta W^{r_2}_{r_f}}{\partial T}\right)dr_2 \label{e55}
\end{eqnarray}
\end {thm}
\textbf{Proof.} Either by subtraction $\left(\Delta H^\std	(T) = \Delta G^\std	(T)  +T \Delta S^\std	(T) \right)$ or the application of the Gibbs-Helmholtz equation of the form$ \frac{\partial}{\partial T}\left(\frac{\Delta G^\std}{T}\right)_{P,\mathbf{n}}=-\frac{\Delta H^\std}{T^2}$, we 
derive from (\ref{e53}, \ref{e54})  the above  result $\bullet$ 

Again, (\ref{e55})~is directly testable if the $P$ function were known. The above equation also explains
why $\xi_{max}\neq \Delta H^\std$ because there are other terms in  (\ref{e55})~  that contributes as well.
\section{\normalsize FREE ENERGY-LIKE RELATIONSHIPS IN KINETICS} \label{sec4}
This section includes a discussion of the Arrhenius rate law and its relation to the equilibrium constant. If we admit that  simple collision theory (SCT) or the quantum version due to Eyring leads to Arrhenius type equations for elementary reactions which determines the rate law then we can state the following:
\begin{thm}\label{th6}
There exists a relationship between the rate constant $k_{r,i}$ and the equilibrium constant $K_e$ given by 
\begin{equation}\label{e64}
\ln k_{r,i}= B(T) + \ln K_e
\end{equation}
where the $B(T)$ term has the form
\begin{equation} \label{e65}
B(T)=\ln\left [A'_i(T)\exp\frac{\overline{W_{rf}}~'}{kT}\right ].
\end{equation}
\end{thm}
\textbf{Proof.} Eqn.(\ref{e53})says that 
\begin{equation}\label{e56}
	\Delta G^\std (T)=\xi_{max}+\overline{W_{rf}}~'
\end{equation}
On the other hand, SCT writes the rate $\nu$ as 
\begin{equation}\label{e57}
\nu = k(\epsilon_r)n^\ast_1n^\ast_2 \,\,\rm{with} \,\,k(\epsilon_r)=\sigma(\epsilon_r)v_r.
\end{equation}
The above expression refers to the rate for the internal kinetic energy $\epsilon_r = \frac{\mu v_r^2}{2}$.where the $n's $ are the densities of the reactants. Integration of $k(\epsilon_r)$ leads to the form of the forward (1) rate constant given by  $k_{r,1}(T)$ \cite [p.99]{hous1} with the rate expressions 
\begin{eqnarray}
	\nu &=&k_{r,1}(T)n^\ast_1n^\ast_2 \,\, \rm{with}\nonumber \\
	k_{r,1}(T)&=& \sigma \bar{v}_{rel}(\mu,T)\exp-\left(\frac{\epsilon^\ast}{kT}\right)\nonumber \\
	 &=& A_1(T)\exp-\left(\frac{\epsilon^\ast}{kT}\right) \label{e58}
\end{eqnarray}

where an Arrhenius form is observed in the last line of (\ref{e58}); $\sigma$ is the maximum impact parameter, $\bar{v}_{rel}(\mu,T)$ is the relative velocity of the particles and $\epsilon^\ast$ is by definition the threshold energy parameter, whereby all reacting molecules must have this energy along the line of centers for reaction to occur. The situation is not altered in Eyring's TST, where $\epsilon^\ast$ goes one step beyond it, in that $\epsilon^\ast$ represents the zero-point energy difference between reactants and activated complex \cite [p.103] {hous1}. Hence, in this model too, $\epsilon^\ast$ is the threshold parameter. For TST theory, the form of the rate constant is 
\begin{equation}\label{e59}
k_{r,2}(T)=\frac{kT}{h}\frac{q^{\neq}}{q_A q_B}\exp(-\frac{\epsilon^\ast}{kT})=A_2(T)\exp(-\frac{\epsilon^{\ast}}{kT})	.
\end{equation}
The $q's$ are the partition functions of the intermediate $\neq$ and reactants $A,B$ \cite[eqn. 3.17,p.105]{hous1}. Both theories are of form $A_i(T)\exp-\left(\frac{\epsilon^{\ast}}{kT}\right)$.
For completeness, we shall augment the forms $k_{r,i}$(which does not alter the form of the standard expressions, and which may be neglected in most studies not concerned with the energies considered here) by a work term $c$ due to the mutual intermolecular forces acting on the reactants, which is a "`real"' term since these forces exist, and which has the form $c=c(n_A,n_B,\mathbf \Omega)$ where $n_A,n_B$ are reactant densities, and $\mathbf {\Omega}$ refers to the physical variables (non-thermodynamical) such as the dielectric constant or the screening parameters . We are focusing on an elementary bi-molecular reactions 	here as an example. We can also apply a steric factor $P$ so that the rate can be written $ 
k_{r,i}=P(T)A_i(T)\exp-\left(\frac{\epsilon^\ast + c}{kT}\right)$.  As $n_A,n_B \rightarrow 0$, we expect $c \rightarrow 0.$ Write $\exp-(\frac{c}{kT})=C'(T)$, so that 
\begin{equation}\label{e60}
	k_{r,i}=P(T)C'(T)A_i(T)\exp-\left(\frac{\epsilon^\ast }{kT}\right).
\end{equation}
Define $A'_i(T)=P(T)C'(T)A_i(T)$ so that 
\begin{equation}\label{e61}
	k_{r,i}=A'_i(T)\exp-\left(\frac{\epsilon^\ast }{kT}\right).
\end{equation}
We interpret the pre-exponential factor $A_i(T)$ to be that which obtains (whenever P is known and specified exactly) whenever the particle density $\rho \rightarrow 0$.
Identifying the $\epsilon^\ast$ threshold with $\xi_{max}$, the threshold energy for coalescence leads to 
\begin{equation}\label{e62}
	\Delta G^\std (T) -\overline{W_{rf}}~'=\xi_{max}=\epsilon^\ast.
\end{equation}
Since the equilibrium constant $K_e$ is given by $-\Delta G^\std (T)=kT\ln K_e$, we are lead to 
\begin{eqnarray}\label{e63}
	k_{r,i}&=&A'_i(T) \exp\left[-\frac{1}{kT}\left\{\Delta G^\std -\overline{W_{rf}}~'\right\}\right]\nonumber \\
	&=& \left\{A'_i(T)\exp\frac{\overline{W_{rf}}~'}{kT}\right\}\cdot K_e .\label{e63}	
\end{eqnarray}
Taking logarithms on both sides of (\ref{e63}) leads to the result $\bullet$. 

Theorem(\ref{th6}) is cast in the standard language of  the "`free energy relationship"' linearly relating for instance the  logarithm of the rate constant with the logarithm of the equilibrium constant \cite[p.961, Sec 27.5]{Ait2} where it is opined that the activation energy is linearly proportional to the standard free energy change for a series of compounds, leading to correlation analysis used in  physical organic studies, fashioned after the Hammett and Taft equations which  are essentially an empirically constructed set of fitting parameters \cite[esp. Chap 3]{hine1}. There are two different ways of approaching Theorem (\ref{th6})  and    (\ref{e64}):
\par
1) From a physical point of view, of little interest to physical organic kineticists, (\ref{e64}) may be viewed as an extention of linear Arrhenius plots, where $\ln k_{r,i}$ is plotted against $1/T^\ast$ to determine the activation energy where $A_i'(T)$ is treated as a constant over the temperature range of the plot whose slope yields the activation energy. In this sense, 
%
the linearity of the log-log plots using (\ref{e64})~is dependent on the stability of the $B(T)$ function of (\ref{e65})~over a large temperature range. Some comments are in order here; that $\ln A'_i(T)$ is experimentally rather  invariant is a fact observed in many   chemical kinetics studies, where the Activation energies are determined routinely by plotting the rate constant against the reciprocal temperature; in these  estimates, $\ln A'_i(T)$ is presumed invariant and accurate values are determined. For the cases considered here, $\overline{W_{rf}}~'$ are linear functions of the temperature, so that division by the temperature factor would lead to a constant expression; actually it would be weakly dependent on temperature due to the $P(r_2,T)$ function, which is expected to have  a slight temperature dependence. Hence, for the \emph{same reaction} conducted at different temperatures, we can predict, subjected to the above assumptions, a somewhat linear relationship between $\ln k_{r,1}(T)$ and  $\ln K_e$. This prediction can be tested in simulations as well, for  the general applicability of (\ref{e64}). 
\par
2) At any one fixed temperature, if the steric $P$ and $C'(T)$ function representing intermolecular work is reasonably constant, then for any two similar reactants A and B, subject to the above conditions, we would expect some forms  of  linear log-log relationships  when  the "`bond"' structure is fairly similar (e.g. similar activation energy  and reaction pathway in the transition state). The correlations would be found if various forms of linear relations are found for   $\overline{W_{rf}}~'$ for   the same reaction with different substituent groups. In view of the fact that there is nothing unique that is specified by Theorem (\ref{th6})~ concerning linear relationships, a great many types of correlations may be expected by the expansion of other  functions that can be related to $\overline{W_{rf}}~'$, such as  $\xi_{max}$ and the masses of the particles constituting the molecule, since these affect the moment of inertia $I$ , and the translational motion. Such an undertaking is a deep study in itself which will not be attempted here, suffice to say that Theorem (\ref{th6})~ can provide a foundation for this undertaking.  


\section{\normalsize CONCLUSION}
Particle reactions moderated or subjected to a temperature field may be treated as a three dimensional (3D) mechanical body which interchanges its internal energy  where these interchanges are due to the interaction of internal forces within the molecule and the thermal reservoir due to change of shape of the composite particle system or "`molecule"'. The analysis above allows for the treatment of single-molecular thermodynamical systems. Advances in technology very recently are gradually bringing about the realization of these entities \cite{li1,li2}. It is possible to predict the standard states of such single-molecular species by measuring its internal motion at equilibrium where the $P(r_2)$ probability distribution function serves as the key in the derivations and this probability function can be determined in principle from single-molecular studies. The linear "`Free Energy"' relationships are  given a firm  theoretical basis because of the derived relationship  between the equilibrium constant and the rate constant for the elementary reaction. A deeper study would derive relationships similar in form to that provided by Hammett and Taft. It is predicted that the height rather than the depth of the potential well would determine the sign of the standard enthalpy and free energy.All these assertions can be tested, and future investigations would be directed toward this end. 
\newline
\appendix \label{a1}

 \appendixname
 \section {\normalsize Time Average Probability Densities From Threshold Energy Values}

The definition of the activation energy is that for $\dot{r}=v_r=0=E_{kin}$ (where $E_{kin}=\frac{1}{2}\mu v_r^2$)$xi_{max}$ is the minimum energy required to form a "`bond"'. During the formation of the bond at the threshold $r_f$, the values of the internal kinetic energy has the range $0<E_{kin},+\infty$.We choose the time increment $\delta t_i$ to be small enough in the grid interval so that the following obtains:
\begin{equation}  \label{e66}
\left|\delta r\right|=\left|r_{i+1}-r_{i}\right|=\beta ,\,\,{\rm where}\,\,\left|\delta t_i v_r\right|<\beta 
\end{equation}
For   $E_{kin}=\alpha$ at $r_f$ when a molecule is formed, the work change $\delta W_{\alpha,i}$ within two adjacent grid sites is 

\begin{equation}  \label{e67}
\delta W_{\alpha,i}=\left( \overline{W_{rf} } (r_{i+1})-\overline{W_{rf} } (r_{i})\right)\neq 0
\end{equation}
at most by property (\ref{e66}); if the bond coordinate difference is $< \beta$,then $\delta W_{\alpha,i}=0$.
We can envisage the intermolecular coordinate $r_i$ hopping by one increment (i.e. from $i$ to $i+i$ or $i-1$).Define an interval    $\delta E$  in energy where $E_\alpha -
\frac{\delta E}{2} < E_\alpha <  E_\alpha   + \frac{\delta E}{2}$ The probability density $P_{r,\alpha}(r_i)$ for a series of individual trials $j$ for the formation of the molecule to its disintegration about energy $E_\alpha$ of interval $\delta E$ about the grid  coordinate $\delta r$ is
\begin{equation}  \label{e68}
P_{r,\alpha}(r_i)=\lim \left.\begin{array}{c} {M\rightarrow \infty}\\
{N'\rightarrow \infty}\\
\delta r \rightarrow 0
\end{array}
\right.\frac{\frac{1}{\delta r}
\sum_{j=1}^M  N_{\alpha,j}(r_i)   }
{\sum_{r=r_0=r_f}^{r=r_{N'}} \sum_{j=1}^{M}N_{\alpha,j}(r_i)}\,\, {\rm with}\,\,\left\{\begin{array}{c}
r_N\rightarrow r_b \\
r_0=r_f.
\end{array} \right.
\end{equation}
Since $\alpha$ varies from $0$ to $+\infty$ we can define the total probability as 

\begin{equation}  \label{e69}
P_{}(r,T)=\lim \left.\begin{array}{c} {M\rightarrow \infty}\\
{\frac{1}{\delta r}\rightarrow \infty}
\end{array}
\right.\frac{
\sum_{\alpha=0}^{\alpha=M\delta \alpha}  P_{r,\alpha}(r_i)   }
{\delta r \sum_{r_i=0}^{r_i=M\delta r} \sum_{\alpha = 0}^{\alpha=M\delta \alpha }P_{r,\alpha}(r_i)}
\end{equation}
and (\ref{e69}) must be equal to (\ref{e21})if the Gibbs' postulate (\ref{ps1}) is correct for a sufficiently dilute system. The shape of the $P(r_2,T)$ is determined by the $E_\alpha$ distribution, together with the internal forces of the bond (molecule) and the mean relaxation time of interaction of the thermal reservoir and the molecule.


  \bibliography{gen1}      
\end {document}